\documentclass[11pt,twoside]{article}
\usepackage{cozumel2005}
\usepackage{epsf}
\usepackage{psfig}
\usepackage{graphicx}
\usepackage{lscape}
\begin{document}
\title{IMF biases and how to correct them}  
\author{J. Ma\'{\i}z Apell\'aniz$^{1,2}$, L. \'Ubeda$^1$, N. R. Walborn$^1$, 
        and E. P. Nelan$^1$}   
\affil{$^1$ Space Telescope Science Institute, 3700 San Martin Drive, Baltimore, 
       MD 21218, U.S.A.}    
\affil{$^2$ Space Telescope Division, European Space Agency, ESTEC, Noordwijk, Netherlands}

\begin{abstract} 
We discuss possible sources of biases in the determination of the initial mass function (IMF) 
introduced by the binning of the data, the uncertainty in the determinations of masses,
and the existence of unresolved multiple systems. Those three effects 
tend to produce IMFs that are flatter than the real one. We analyze the importance of each
effect and suggest techniques that minimize or eliminate the biases. We also report the 
detection of the first astrometric binary system composed of two very-early O-type stars, 
HD~93129~A.
\end{abstract}


\section{Introduction}

$\,\!$ \indent In order to obtain the mass function of a stellar population from photometric data one starts by placing 
the stars in a color-magnitude or in a theoretical (temperature-luminosity) HR diagram 
along with the evolutionary tracks and the corresponding
isochrones. For a simple population one can then find the appropriate isochrone and obtain the masses for each star. For a complex 
star formation history, one has to obtain the full transformation from temperature and luminosity (or color and magnitude) to mass 
and age. The number of stars as a function of mass (corrected for age effects if necessary) can then be used to obtain the initial
mass function (or IMF).

	A number of problems and biases can arise along the way to obtaining the true IMF due to the oversimplification of the 
assumptions (e.g. using isochrone fitting for a population with an age spread), the existing intrinsic degeneracies (e.g. 
the complicated topology of evolutionary tracks, metallicity effects, and rotation can make two stars of different masses and ages 
have the same temperature and luminosity), as well as for other reasons. In this work
we will analyze three sources of systematic effects: the numerical bias introduced by the use of constant-size bins for the 
fitting of the IMF, the ``mass diffusion'' from low to high masses due to photometric uncertainties, and the existence of unresolved
multiples. Those three effects go in the same direction of making the measured IMF flatter than the real one and can affect
different samples to different degrees, thus introducing the possibility of yielding a dispersion of measured IMF values where only
a single one exists in reality. 

\section{Binning biases}

$\,\!$\indent The first type of bias we will discuss is purely numerical and affects not only IMF determinations but the fitting of any
function to binned data that follows Poisson or multinomial statistics \citep{Wheaetal95,Lucy00} and was analyzed by
\citet{MaizUbed05}. Suppose we have measured the masses ($m$) for a set of stars to which we want to fit a power law of the form:

\begin{equation}
\frac{dn}{dm} = A \cdot m^{\gamma},
\label{imf1}
\end{equation}

\noindent where $dn$ is the number of stars with mass in the interval $m$ to $m+dm$. Integrating both sides of the equation and
taking logarithms we arrive at the expression:

\begin{equation}
\log_{10} N_i =
\log_{10} \left( \frac{A}{\gamma + 1}  \cdot \left[   \left( x_i + \frac{\Delta m_i}{2} \right)^{\gamma + 1}-  
\left( x_i-\frac{\Delta m_i}{2} \right)^{\gamma + 1} \right]  \right),
\label{imf2}
\end{equation}

\noindent where $x_i$ is the mass at the center of an interval of width $\Delta m_i$ that contains $N_i$ stars. Equation~\ref{imf2}
is the expression that can be used to derive the IMF slope, $\gamma$, by measuring the number of stars in each bin and fitting the
data, which can be done by minimizing a $\chi^2$ statistic and deriving the associated uncertainties. Since $N_i$ follows a 
binomial distribution (with $N$ being the total number of stars summed over all bins), the associated weight for bin $i$ for the 
$\chi^2$ fit is given by:

\begin{equation}
w_i = \frac{N_iN}{(N-N_i)(\log_{10}e)^2}.
\label{weights}
\end{equation}

	Two warnings should be given here. The first one is that for two bins $i$ and $j$, $N_i$ and $N_j$ are correlated because
they are both part of a joint multinomial distribution (see e.g. \citealt{Lucy00}). The second and most important one is 
that the value of $N_i$ that should be strictly used in Eq.~\ref{weights} is the one calculated from the fit, not the one measured
from the data \citep{Wheaetal95}. Otherwise, one runs into the possibility of introducing biases in the measurement of $\gamma$ unless
some precautions are taken because of the differences between the real and the assumed weights assigned to bins with a small number
of stars in them. 

\begin{figure}
\centerline{\includegraphics*[width=0.475\linewidth]{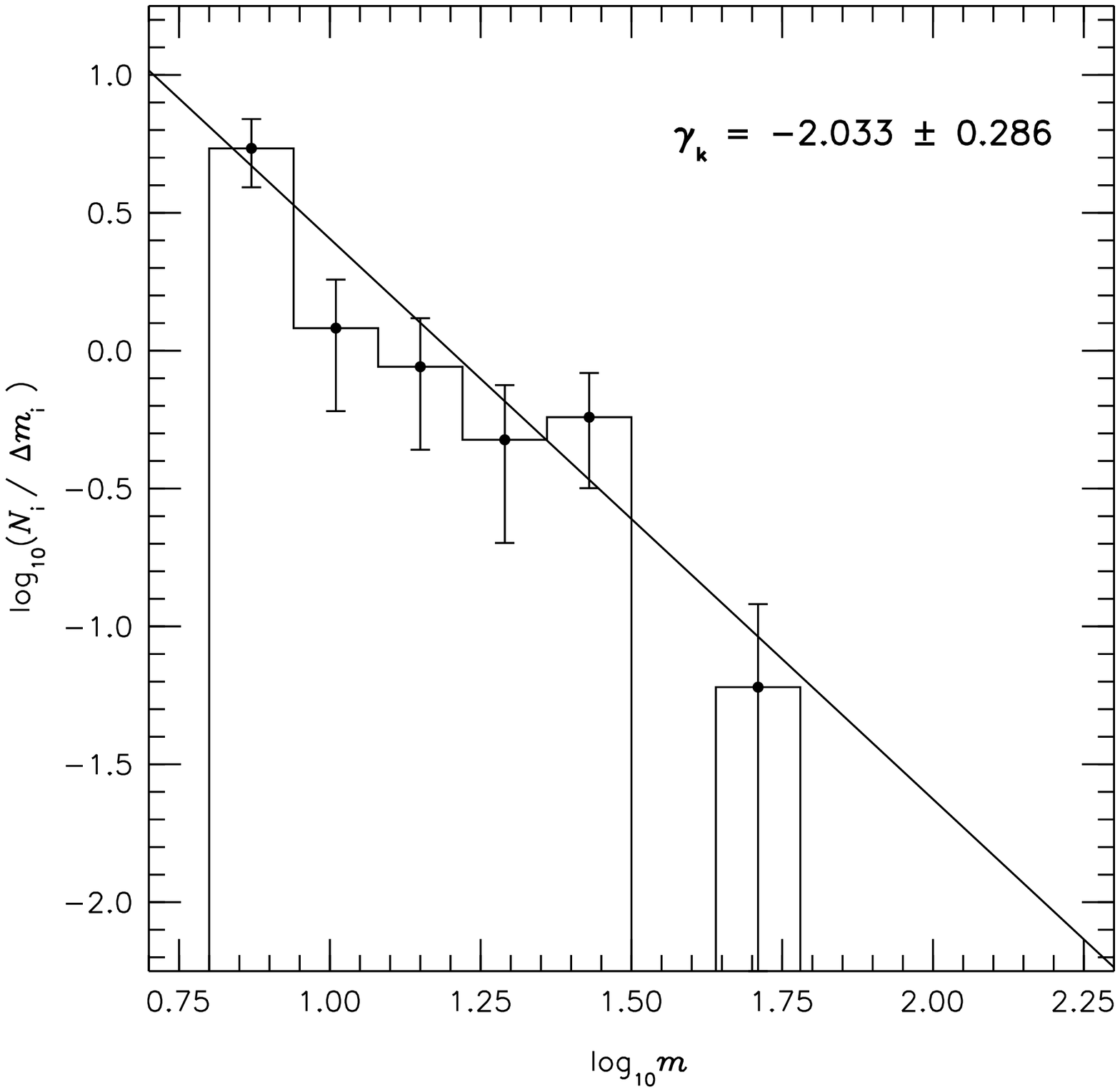} \
            \includegraphics*[width=0.475\linewidth]{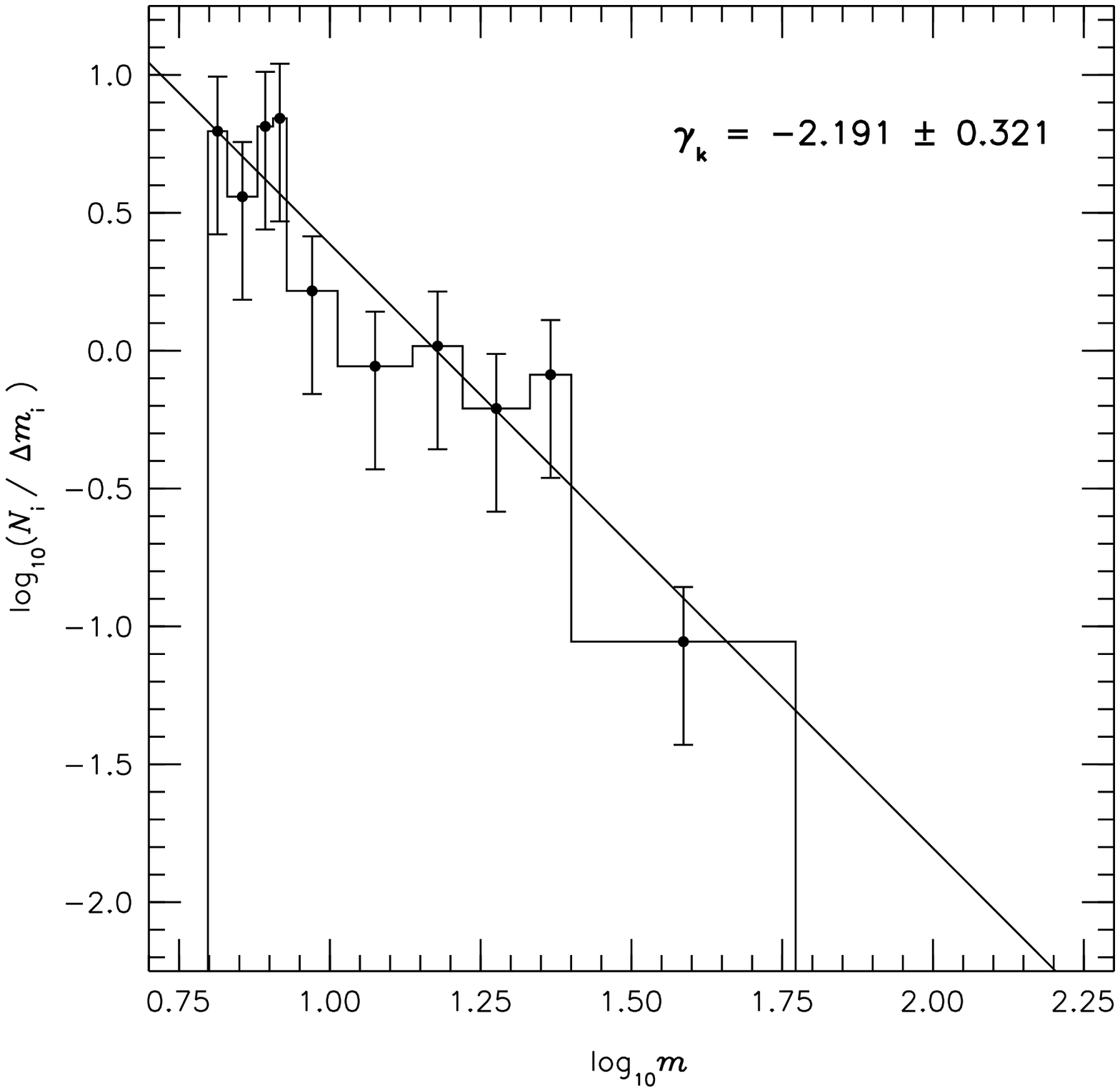}}
\caption{Comparison between the data and the fitted functions for one of the realizations with
30 stars and 10 bins for the two numerical experiments described in the text. The left panel 
corresponds to the first experiment (constant bin size, fixed lower and upper mass limits) while the
right panel corresponds to the second experiment (variable bin size with object equipartition, 
lower and upper mass limits determined from the data). Note the differences in the size of the error bars
between the two experiments and that the left panel plot includes four bins with zero stars.}
\label{comparehisto}
\end{figure}

	Using the fit $N_i$ instead of the data $N_i$ for the weights requires iterating and, therefore, complicates the IMF calculation. 
An alternative strategy was analyzed by \citet{MaizUbed05}: rather than trying to find the right weights for each bin, one can
select the bins in such a way that they all have similar weights and, therefore, the obtained value of $\gamma$ is
nearly independent of the weights themselves. In this manner, the weights derived from the data should give very similar results to
the weights derived from the fit, hence eliminating the need for an iterative procedure.
This strategy can be implemented by using bins with object equipartition, i.e.  selecting the bins in such a way that they all have the 
same number of stars in them. A graphical example of the difference between constant bin-size (the
standard approach to fitting binned data) and variable bin-size with object equipartition can be seen in Fig.~\ref{comparehisto}.

\begin{table}
\caption{Normalized biases for the first experiment (constant bin size) for 3, 5, 10, 30, and 50 bins.}
\centerline{\begin{tabular}{rrrrrr}
 & & & & & \\
\hline
stars & \multicolumn{5}{c}{$b$} \\
 & \multicolumn{1}{c}{3} & \multicolumn{1}{c}{5} & \multicolumn{1}{c}{10} & \multicolumn{1}{c}{30} & \multicolumn{1}{c}{50} \\
\hline
   30   & 0.376  &  0.655  &  1.181  &  2.393  &  2.986  \\     
   100  & 0.176  &  0.376  &  0.772  &  2.058  &  2.988  \\     
   300  & 0.121  &  0.224  &  0.430  &  1.384  &  2.200  \\     
   1000 & 0.151  &  0.163  &  0.260  &  0.766  &  1.275  \\     
\hline
\end{tabular}}
\label{constantbin}
\end{table}

\begin{table}
\caption{Normalized biases for the second experiment (bin equipartition) for 3, 5, 10, 30, and 50 bins.}
\centerline{\begin{tabular}{rrrrrr}
 & & & & & \\
\hline
stars & \multicolumn{5}{c}{$b$} \\
 & \multicolumn{1}{c}{3} & \multicolumn{1}{c}{5} & \multicolumn{1}{c}{10} & \multicolumn{1}{c}{30} & \multicolumn{1}{c}{50} \\
\hline
   30   & 0.110  &  0.134  &  0.180  &  0.264  & \ldots  \\     
   100  & 0.047  &  0.046  &  0.071  &  0.143  &  0.161  \\     
   300  & 0.053  &  0.066  &  0.065  &  0.065  &  0.080  \\     
   1000 & 0.044  &  0.079  &  0.079  &  0.073  &  0.086  \\     
\hline
\end{tabular}}
\label{variablebin}
\end{table}

\begin{figure}
\centerline{\includegraphics*[width=0.468\linewidth,bb=28 14 566 566]{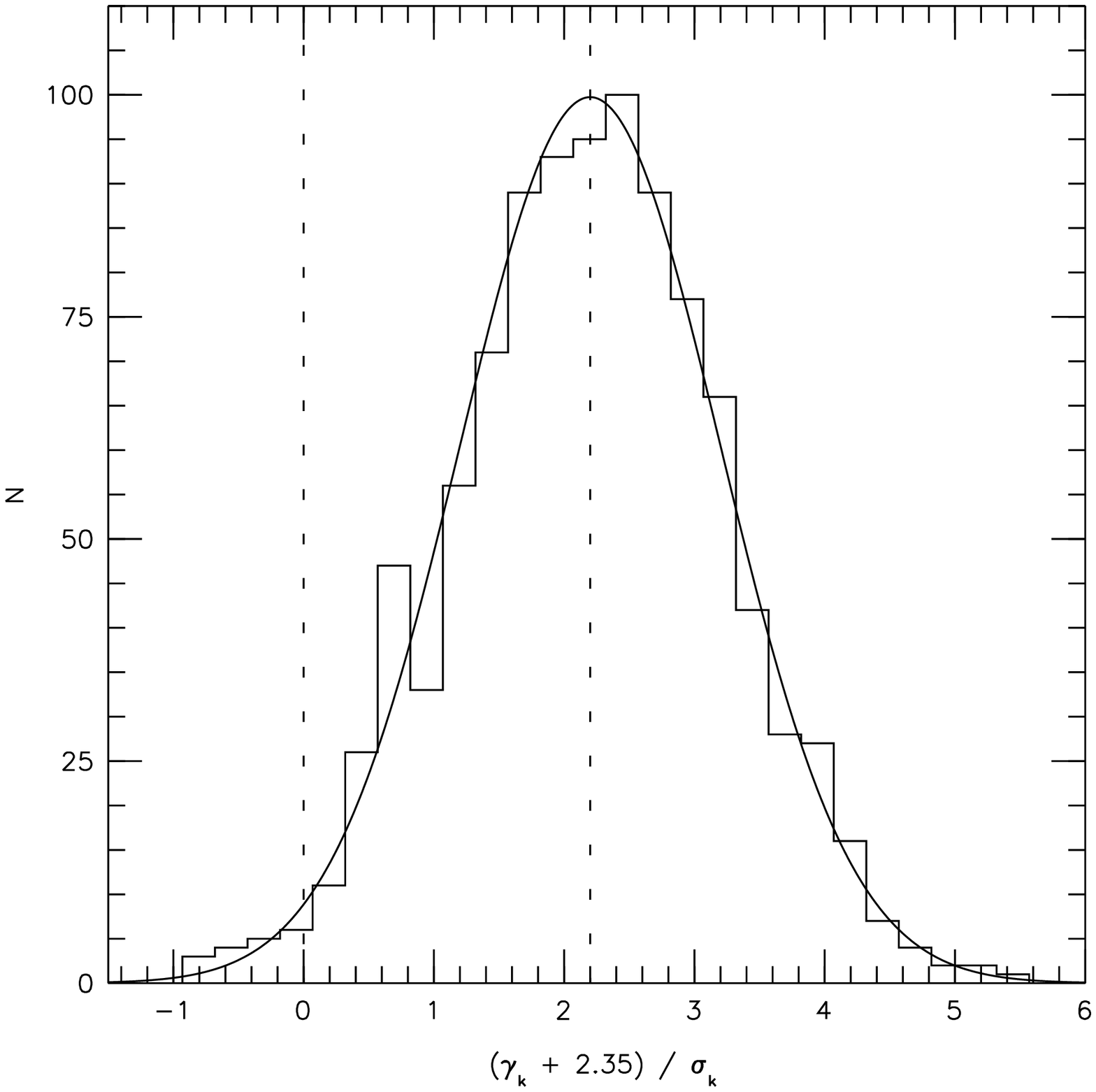} \
            \includegraphics*[width=0.482\linewidth,bb=28 20 566 566]{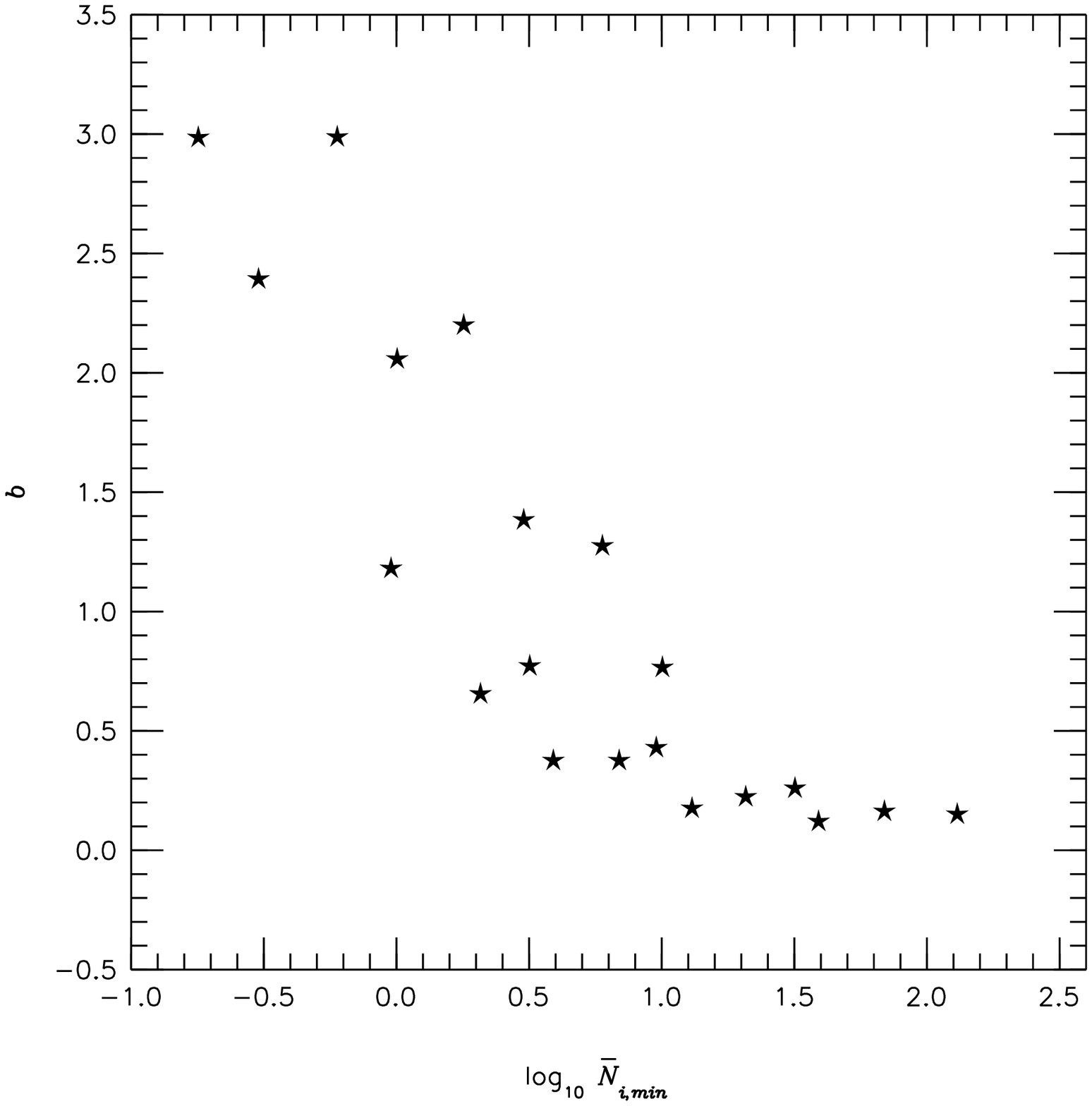}}
\caption{(left) Histogram with the distribution of $(\gamma_k+2.35)/\sigma_k$ for the 1000 realizations of 
the first experiment with 300 stars and 50 bins. A Gaussian distribution with mean $b=2.200$ and 
dispersion of 1.0 is also plotted for comparison. The vertical lines mark the position of 0 and of $b$.
(right) Bias as a function of $\overline{N}_{i,{\rm min}}$ for the first experiment. Note that
$\overline{N}_{i,{\rm min}}$ can be smaller than 1 because it is a property derived from the
parent distribution.}
\label{exp1}
\end{figure}

\begin{figure}
\centerline{\includegraphics*[width=0.468\linewidth,bb=28 14 566 566]{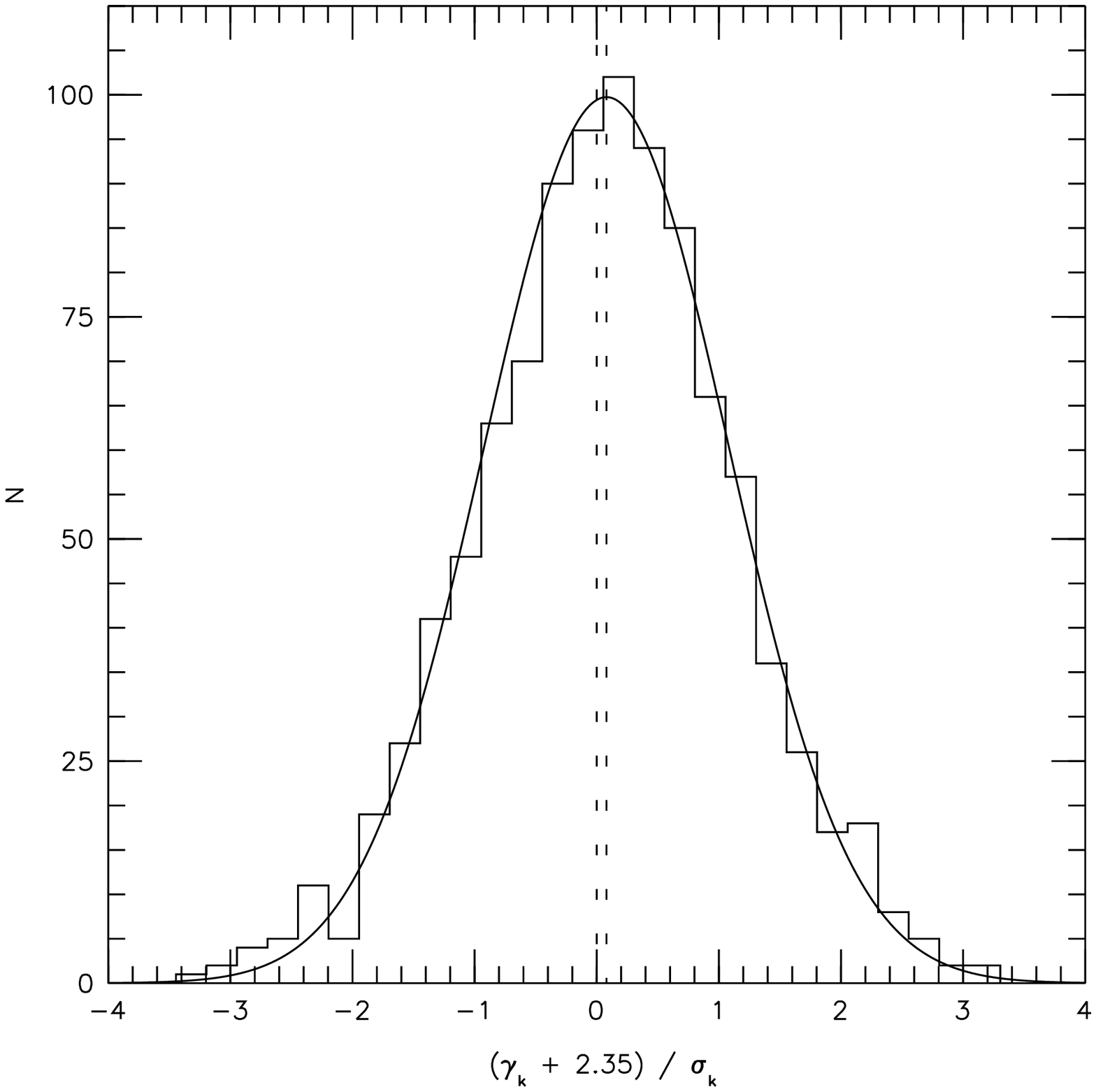} \
            \includegraphics*[width=0.482\linewidth,bb=28 20 566 566]{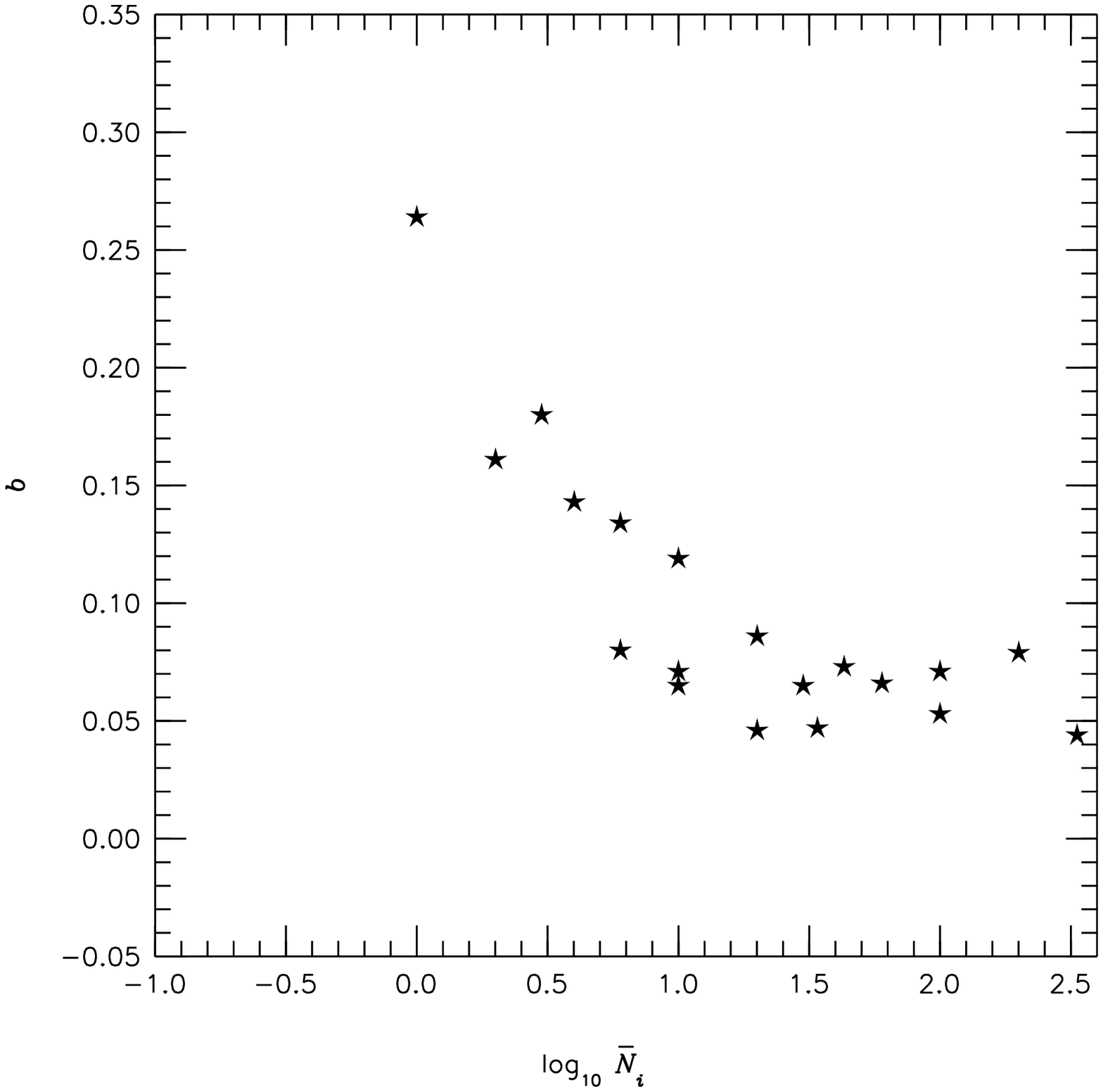}}
\caption{(left) Histogram with the distribution of $(\gamma_k+2.35)/\sigma_k$ for the 1000 realizations of 
the second experiment with 300 stars and 50 bins. A Gaussian distribution with mean $b=0.090$ and 
dispersion of 1.0 is also plotted for comparison. The vertical lines mark the position of 0 and of $b$.
(right) Bias as a function of $\overline{N}_{i}$ for the second experiment. Note that the vertical
scale for the plot is 1/10 of that of the right panel in Fig.~\ref{exp1}.}
\label{exp2}
\end{figure}

	In order to test whether variable bins provide a significant improvement over constant bins, \citet{MaizUbed05} 
designed a series of Montecarlo numerical experiments in which they also analyzed the importance of the choice 
of the lower and upper mass limits. The reader is referred to that article for details; here we provide a summary in 
Tables~\ref{constantbin}~and~\ref{variablebin} and in Figs.~\ref{exp1}~and~\ref{exp2} of two of the experiments. In the
first experiment, they generated 1000 realizations from a distribution with Salpeter ($\gamma=-2.35$) slope using either
$N$ = 30, 100, 300, or 1000 stars. Each realization was then binned into 3, 5, 10, 30, and 50 constant-size bins and a power law
was fitted using $\chi^2$ minimization. Finally, for the 1000 realizations of each $N$ + number of bins combinations
the normalized bias $b$ was computed using the definition:

\begin{equation}
b = \frac{1}{1000}\sum_{k=1}^{1000} \frac{\gamma_k+2.35}{\sigma_k},
\end{equation}

\noindent where $\sigma_k$ is the uncertainty in $\gamma_k$ derived from $\chi^2$ minimization and 
the $k$ index is used to denote the realization number. $b$ is a sensible choice to judge the 
existence of biases. If $|b|\ll 1$, then the fitting method will be unbiased because it will yield values that will be larger 
than the real one on $\approx$50\% of the occasions and smaller on another $\approx50$\%. If, on the other hand, $|b|\sim 1$ or 
larger, a significant bias will exist. 

	As seen in Table~\ref{constantbin}, significant biases exist for most $N$ + number of bins combinations when using constant
bins. In the right panel of Fig.~\ref{exp1} we see that $b$ is a strong function of $\overline{N}_{i,{\rm min}}$, the mean $N_i$ 
in the bin with the lowest number of counts (which for constant-size bins and a Salpeter power law will be the rightmost one). Note
that at least 10 stars in the rightmost bin are required for this method to yield small biases, thus making its use impractical for most
applications. Also note that the bias always points in the direction of making the measured IMF flatter than what it really is and
that, for a fixed number of bins, the effect is larger when there are fewer stars.

	The second experiment (third in the numbering of \citealt{MaizUbed05}) was a repetition of the first one using bin
equipartition instead of constant bins. As a comparison of Tables~\ref{constantbin}~and~\ref{variablebin} shows, the biases for the 
second experiment are much smaller (even for the extreme case of dividing 30 stars into 30 one-star bins, $b$ is only 0.264). The 
effect is also perceptible in the right panel of Fig.~\ref{exp2}, where the vertical scale is a factor of 10 smaller than in the
equivalent panel of Fig.~\ref{exp1}. Furthermore, the left panel of Fig~\ref{exp1} shows that the distribution of 
$(\gamma_k+2.35)/\sigma_k$ is well approximated by a Gaussian with a dispersion of 1.0, indicating that the random uncertainties 
derived from $\chi^2$ minimization can also be trusted. 

	We conclude that the bin equipartition strategy proposed by \citet{MaizUbed05} for the fitting of power laws
with Salpeter slopes yields results that (a) are nearly bias-free and (b) produce correct uncertainty 
estimates. On the other hand, the standard uniform-size binning
introduces biases that are dependent on the number of stars per bin. The power of the equipartition technique
extends to small samples, since it is possible to obtain accurate values
with reasonable precisions for the IMF slope even when as few as 30 stars are available for analysis.

	We would also like to point out that, given the purely numerical nature of the analysis, these
results could be extended to other similar problems. For example, the mass function for young stellar
clusters can be rather well approximated by a power law with a slope of $-2.0$ (see e.g. 
\citealt{FallZhan01}), which is quite close to $-2.35$, so the same type of biases should be present 
there as well. In general, we recommend that biases be evaluated for any function fitted to binned data
through $\chi^2$ minimization by means of specific numerical experiments analogous to the ones in this 
article.

\section{Mass diffusion}

$\,\!$\indent The second type of bias we will discuss is caused by photometric uncertainties and detection limits. When one tries to
use photometric data to derive statistical properties of a stellar population, one finds that corrections for the undetected stars must 
be included in the calculation because it is easier to detect bright stars than dim ones. Such an incompleteness 
correction is usually handled by crowded-field photometry packages such as DAOphot \citep{Stet87} or HSTphot \citep{Dolp00a} by
doing experiments in which the code is run with artificial stars added and the percentage of recovered objects as a function of
magnitude and color is calculated. The correction is then applied to the observed stellar statistics. 

	Obviously, ignoring an incompleteness correction can result in a large bias in the measurement of the IMF and this
well-known fact is taken into consideration in modern works on the subject. However, a related bias which is more subtle is not
always taken into account. Suppose that we observe several times a star that has a real magnitude $m$. Due to Poisson, detector, 
and background noise, in some of our observations we will measure a magnitude $m^\prime > m$ and in others we will measure 
$m^\prime < m$. If our detector is well calibrated, the first circumstance will happen 50\% of the time and the second one the
remaining 50\%. From our analysis of the detector we should be capable of estimating from a single measurement of the star an
uncertainty $\sigma_m$ in such a way that $m^\prime\pm\sigma_m$ behaves in an approximately Gaussian way,
e.g. the single-measurement values $m^\prime$ will be within $m-\sigma_m$ and $m+\sigma_m$ for approximately 2/3 of
the sample and outside ($m-2\sigma_m,m+2\sigma_m$) for approximately 5\% of the sample. Now, for most of the stellar mass range the
IMF has a negative slope, meaning that there are more low-mass (dim) stars than high-mass (bright) ones. Therefore, if we measure a
star to have magnitude $m^\prime$, there should be a higher probability that its real magnitude $m$ is dimmer than $m^\prime$ than
that it is brighter (i.e. there are more dim stars disguised as bright ones at a given measured magnitude than bright stars
disguised as dim ones) because the underlying real luminosity distribution provides more dim stars to start with. We will call this
effect mass diffusion because it acts in a manner analogous to a diffusion process, smoothing a gradient by shifting objects from
where they are more abundant to where they are more scarce.

\begin{figure}
\centerline{\includegraphics*[width=\linewidth]{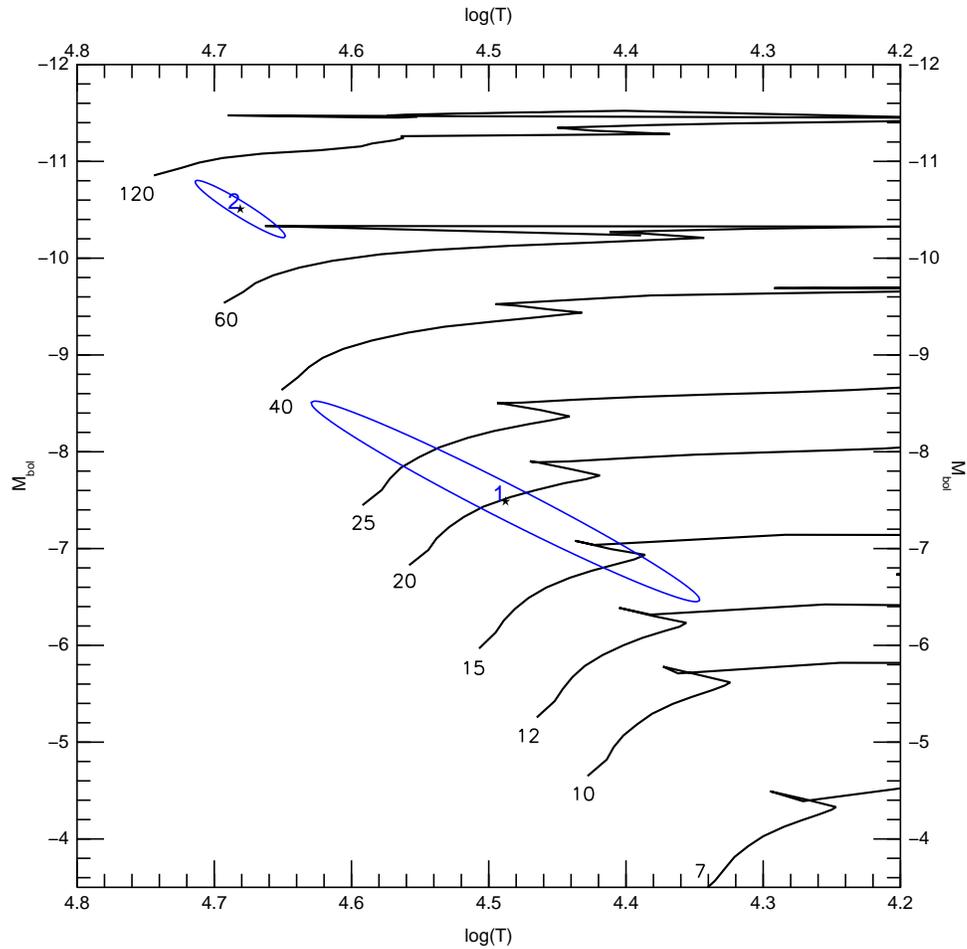}}
\caption{Converting from photometry-derived temperatures and luminosities to masses. This plot shows the temperatures and
luminosities for two stars in NGC 4214 derived from HST/WFPC2 F170W+F336W+F555W+F814W photometry and their associated uncertainty
ellipses. Also shown are evolutionary tracks from \citet{MaedMeyn01}.}
\label{massdif}
\end{figure}

	Computing the correction required to eliminate mass-diffusion effects from an IMF is not straightforward because two
intermediate steps are required. First, one has to translate uncertainties in the measured magnitudes and colors into uncertainties
in temperature and luminosity (or bolometric magnitudes). Second, the uncertainties in the theoretical HR diagram have to be
converted into uncertainties in mass. The first step involves applying extinction and bolometric corrections and taking into 
consideration the correlations between them and the measured magnitudes. An example of such effects is shown in Fig.~\ref{massdif},
where we have plotted the temperatures and bolometric magnitudes of two stars in NGC 4214 derived from multiband HST/WFPC2 stellar
photometry \citep{Ubedetal05}. A strong correlation is observed between temperatures and bolometric magnitudes: most of this
correlation is caused by the strong temperature dependence of the bolometric correction. The data plotted in Fig.~\ref{massdif} 
was calculated using CHORIZOS \citep{Maiz04c}, a code specifically designed for the task of transforming from measured magnitudes to
physical properties such as temperature, age, or extinction. Note that the lower uncertainty ellipse is larger than the upper one:
most of this effect is caused by the higher uncertainties in the measured magnitudes of dim stars compared to bright ones. 

	The second step (calculating the uncertainties for the masses) can be achieved by producing a point-to-point coordinate conversion 
between temperature + luminosity and mass + age (with the caveats about one-to-one correspondence previously mentioned), 
generating a distribution of temperatures and luminosities for each star according to its uncertainty ellipse,
transforming those values into masses and ages using the conversion above, and calculating the mean and standard deviation of the
derived mass distribution for each star. Those values can then be used as the mass ($M_i$) and its uncertainty ($\sigma_{M_i}$) for 
each star.

\begin{figure}
\centerline{\includegraphics*[width=\linewidth]{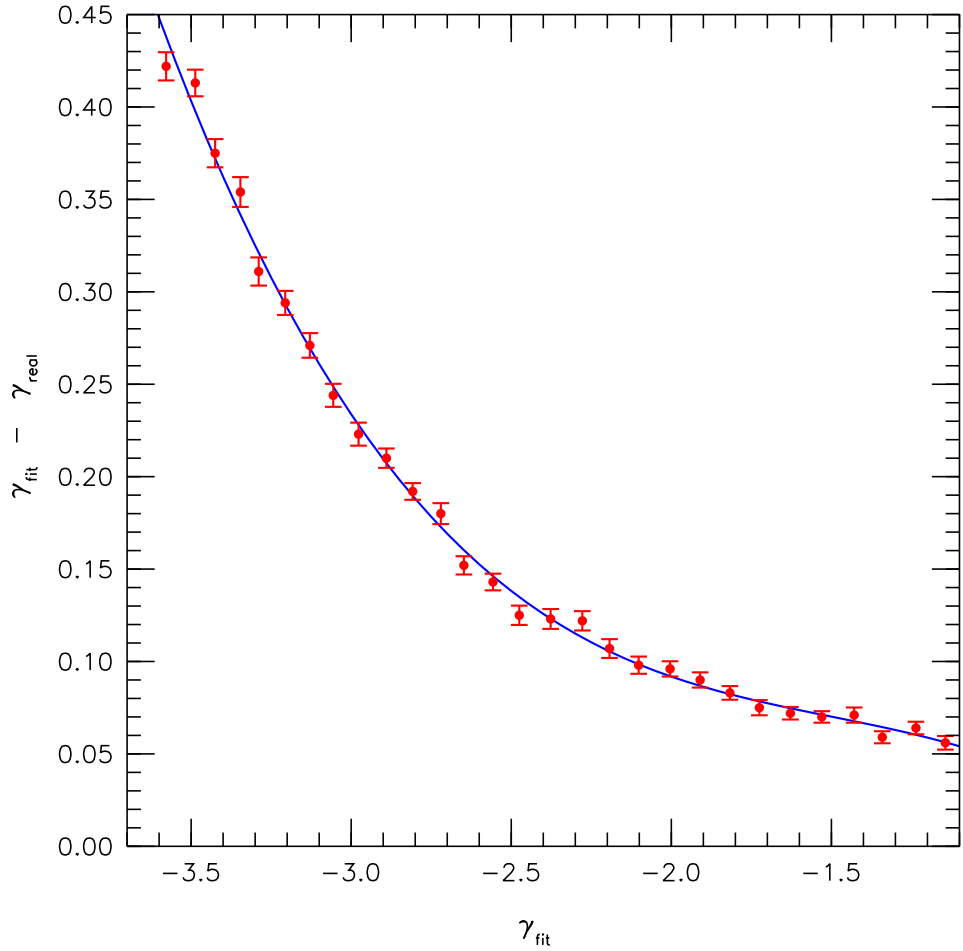}}
\caption{Correcting mass-diffusion biases. The plot shows the difference between the fitted and the real slopes of the IMF as a 
function of the fitted value for the artificial IMF realizations described in the text. The individual data points are the results of
the fit to the realizations and the continuous line is a polynomial fit. The $\sigma_M(M)$ and the total number of stars used are
those corresponding to the NGC 4214 data of \citet{Ubedetal05}.}
\label{massdifcor}
\end{figure}

	Once the masses and corresponding uncertainties have been computed for each star, one can calculate the correction due to mass
diffusion in the following way. First, a function $\sigma_M(M)$ is computed from the values for the individual stars 
($M_i$, $\sigma_{M_i}$) by fitting a simple function such as a parabola. Then, one generates a series of realizations of the IMF with
different values of the slope (which we will call $\gamma_{\rm real}$) which are then smoothed using a Gaussian kernel with 
$\sigma_M(M)$. The output is fitted using $\chi^2$ minimization and the corresponding fitted value of the slope (which we will call
$\gamma_{\rm fit}$) is obtained. Finally, as shown in Fig.~\ref{massdifcor}, a polynomial is fit to 
$\gamma_{\rm fit}-\gamma_{\rm real}$ as a function of $\gamma_{\rm fit}$ which is the correction that needs to be applied. Note that
since $\sigma_M(M)$ is data-dependent, an individual correction has to be applied to each specific observation.

	We show in Fig.~\ref{massdifcor} the correction for the specific case of the NGC 4214 data previously mentioned. The magnitude
of the correction can be taken to be typical for HST photometry beyond the Magellanic Clouds obtained with a few orbits of exposure
time. The bias produced by not applying the correction is in the same direction as the one caused by constant-size bins: the measured
IMF appears to be flatter than the real one. 

	We recommend that the correction described here be applied to the calculation of IMFs in general. However, we should point out
that, ideally, one would like the correction to be as small as possible. One (obvious) way to achieve this is to obtain photometry with better 
S/N ratio.  Another one is to use as many filters as possible to adequately characterize the temperature (and possibly gravity and 
metallicity) of each star and to adequately
correct for extinction and then to process the data using a code like CHORIZOS, instead on relying on conversions from single-color +
magnitude diagrams to temperature + luminosity equivalents.

\section{Multiplicity}

$\,\!$\indent The third type of bias that will be discussed here is that caused by multiplicity. It has been known for a long time that
a large fraction of stars are located in multiple systems. For example, \citet{Kouwetal05} measured that at least 61\% of the stars in
the Scorpius-Centaurus OB association are in a multiple system (the number could actually be higher due to incompleteness and selection
effects). Interestingly, the single-star fraction decreases for early spectral types: those authors measured that 41\% of the systems
in which the primary is a B4-B9 are single but the fraction decreases to 12\% if the primary is a B1-B3. The large multiplicity at
the high-mass end of the stellar mass spectrum is not a new result: \citet{Masoetal98} measured that at least 75\% of the stellar systems
with O stars in clusters or associations are multiple. Furthermore, those authors recognized that with the current instrumentation
capabilities there is still enough discovery space between visual and spectroscopic binaries to allow for basically all O stars in
clusters and associations to be part of multiple systems.

	The existence of unresolved binaries artificially flattens the IMF due to a combination of two effects. First, an unresolved binary
of any mass is shifted from a lower mass to a higher mass in a mass histogram. 
Second, the flattening should be further enhanced by the apparent increase
of multiplicity with mass in the range between a few and several tens of solar masses, which is the range for which Salpeter slopes are
reported by most authors (\citealt{Chab03} and references therein). 

	The flattening due to unresolved binaries has been estimated by \citet{Krou01} to produce a change in $\gamma$ between 0.0 and 1.3 
for low-mass stars and brown dwarfs. Given the large multiplicity fractions observed for OB stars, 
the effect must also be significant for intermediate- and high-mass stars.
At the highest end of the stellar mass spectrum, the problem of the calculation of the IMF slope is coupled with another one: is there an
upper limit for stellar masses or is the highest mass in a cluster simply determined by statistical sampling in a quasi-Salpeter power law
that extends to infinite masses? A few years ago there was a large discrepancy between the highest mass measured from orbital motion
(the most reliable method of measuring masses), which yielded values around 60 M$_\odot$, and the masses measured from photometry and spectral
classification in R136 by \citet{MassHunt98}. Those authors measured stellar masses in the range 120-150 M$_\odot$ and claimed that the data
were compatible with a Salpeter IMF that extended beyond there. That gap has been recently narrowed in both directions. On the one hand, 
the 60 M$_\odot$ barrier for spectroscopic-binary masses has been broken and the current heavyweight champion, WR 20a, lies around 
80 M$_\odot$ \citep{Rauwetal04,Bonaetal04}. On the other hand, new statistical analyses indicate that there appears to be an upper 
mass limit somewhere in the 120-200 M$_\odot$ range \citep{WeidKrou04,OeyClar05,Fige05}.

	In order to study the effect of unresolved binaries on the slope of the IMF at its upper-mass end and to obtain a better
constraint on the stellar upper mass limit, we are currently engaged in an HST GO program (10602) which is a continuation of a previous
one (10205). We are obtaining multi-filter imaging of all known Galactic O2/O3/O3.5 stars with the High Resolution Channel (HRC) on the
Advanced Camera for Surveys (ACS). The HRC has a pixel size of
0\farcs027 and its PSF is well sampled and very stable across the detector \citep{AndeKing04}. Furthermore, its geometric distortion is 
very well characterized, allowing for a relative astrometric precision of 0.005 pixels for very bright stars. 

\begin{figure}
\centerline{\includegraphics*[width=\linewidth]{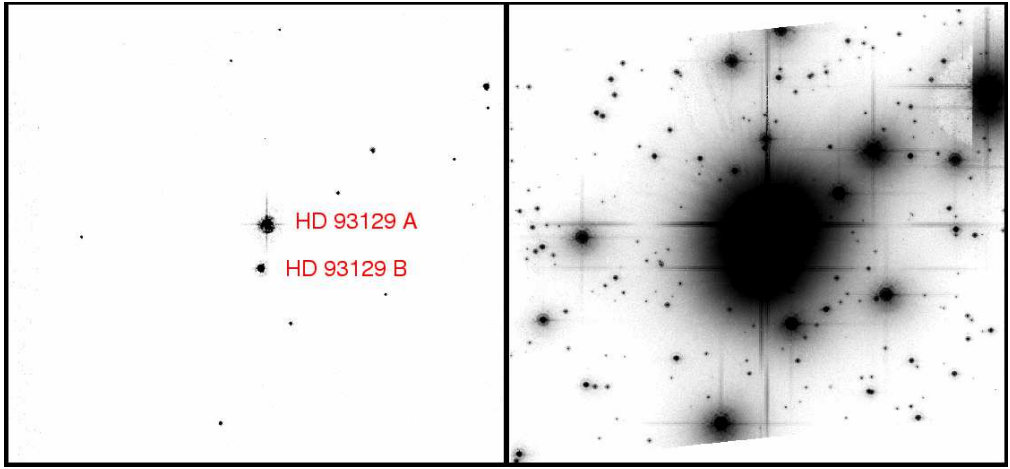}}
\caption{HRC images of the core of Trumpler 14. The left panel shows the F435W ($B$) image and the right panel shows
the F850LP ($z$) image. The field size is $31.0\arcsec\times28.6\arcsec$. Top is 23.2 degrees West of North. Geometric 
distortion has been removed from the images but some cosmetic artifacts are still present in the F850LP case.}
\label{trumpler14}
\end{figure}

	We present here our first results on the core of Trumpler 14 (Fig.~\ref{trumpler14}), 
a young cluster in the Carina Nebula Association that contains at 
least three very-early O-type stars, including HD 93129 A, of spectral type O2 If*, the closest known O2 star 
\citep{Walbetal02b,Maizetal04b}, and HD 93129 B, which is an O3.5 V((f+)). Those two stars are separated by 2\farcs7 and appeared to be 
single in ground-based speckle interferometry \citep{Masoetal98}. Recently, however, \citet{Nelaetal04} were able to split HD 93129 A 
into two components using the Fine Guidance Sensor (FGS) on HST. 
They obtained a separation of 55 $\pm$ 3 mas at a position angle of 356 $\pm$ 4 degrees (measured 
from N towards E) and a magnitude difference of 0.90 $\pm$ 0.05 in the visible. In the same data HD 93129 B is unresolved.

\begin{figure}
\centerline{\includegraphics*[width=0.32\linewidth]{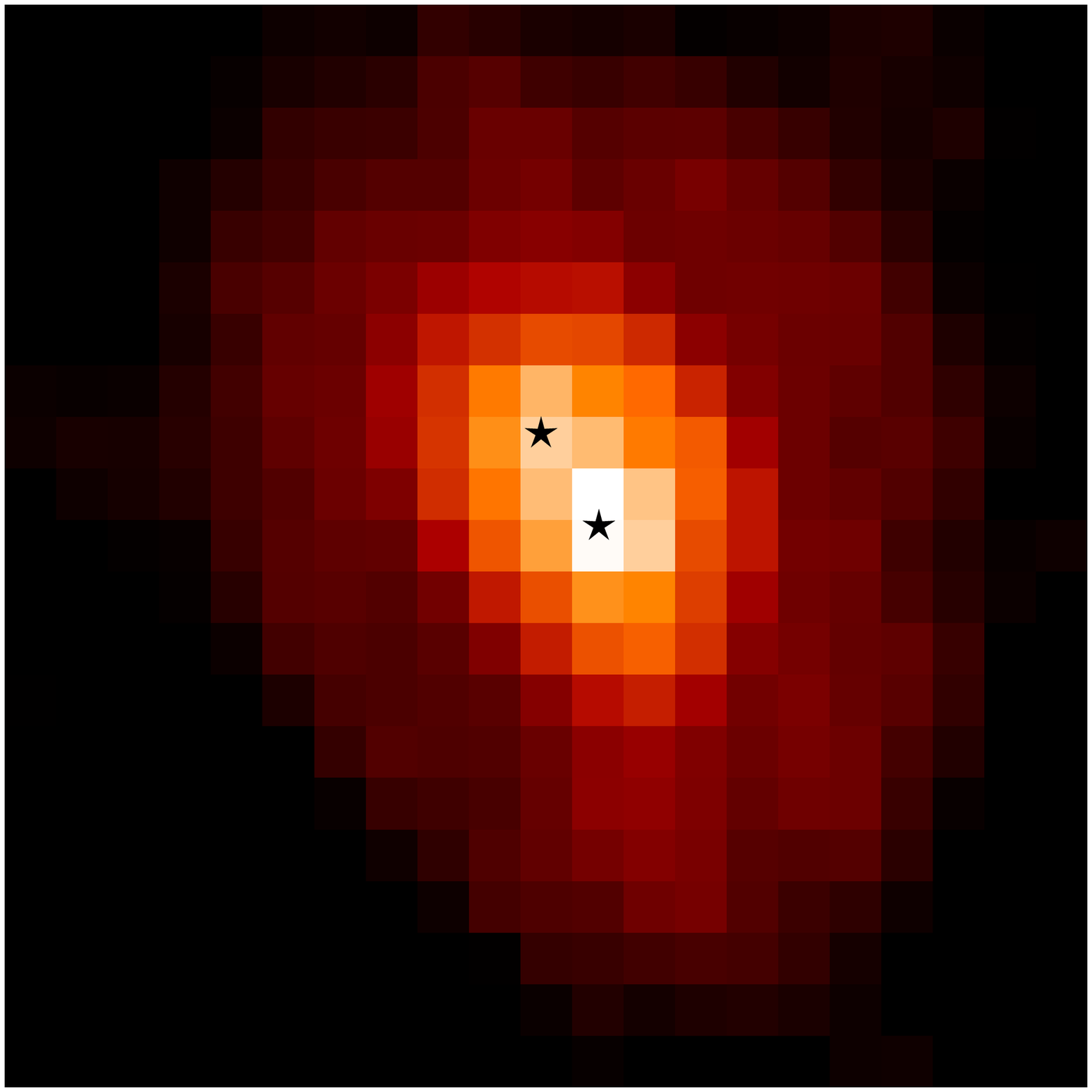} \
            \includegraphics*[width=0.32\linewidth]{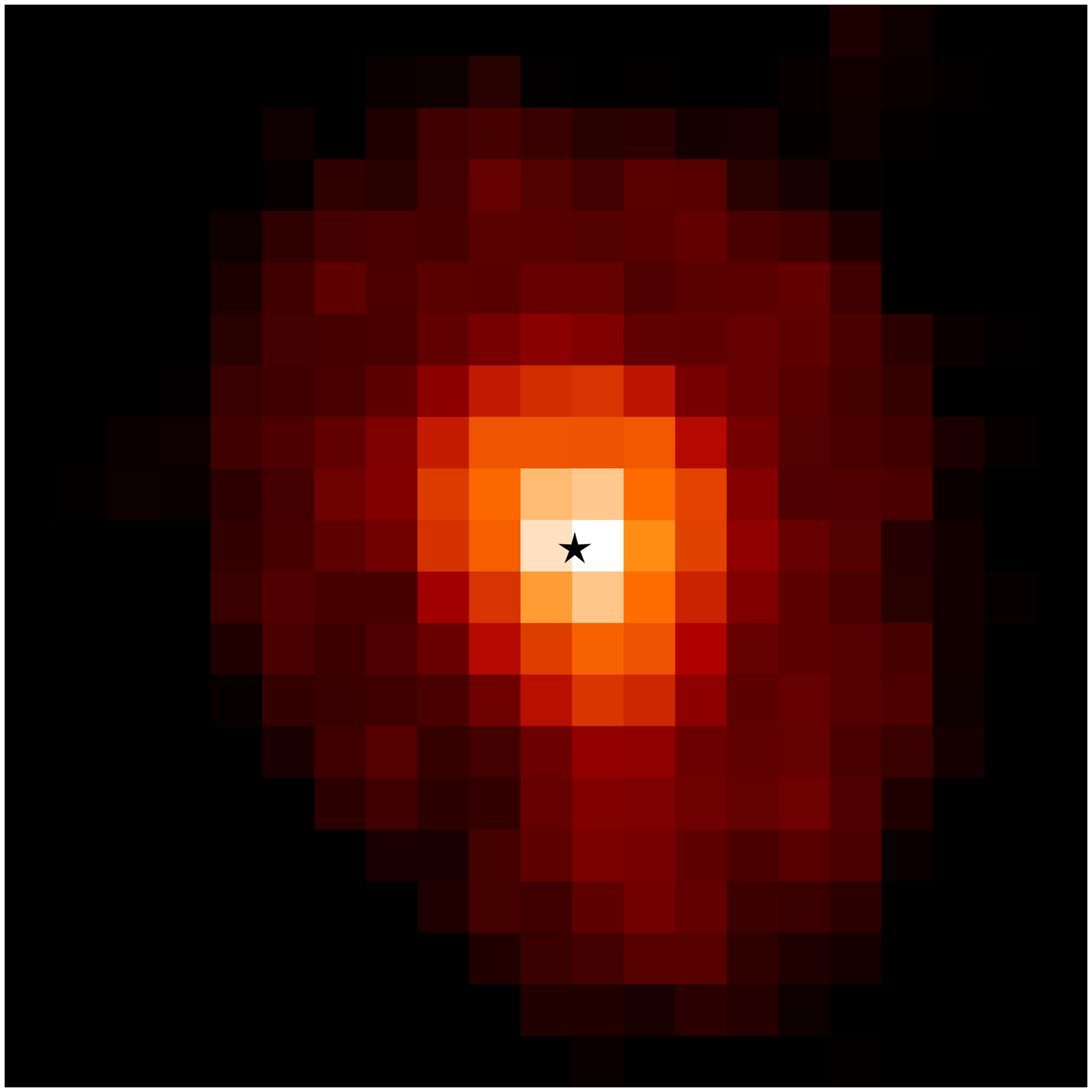} \
            \includegraphics*[width=0.32\linewidth]{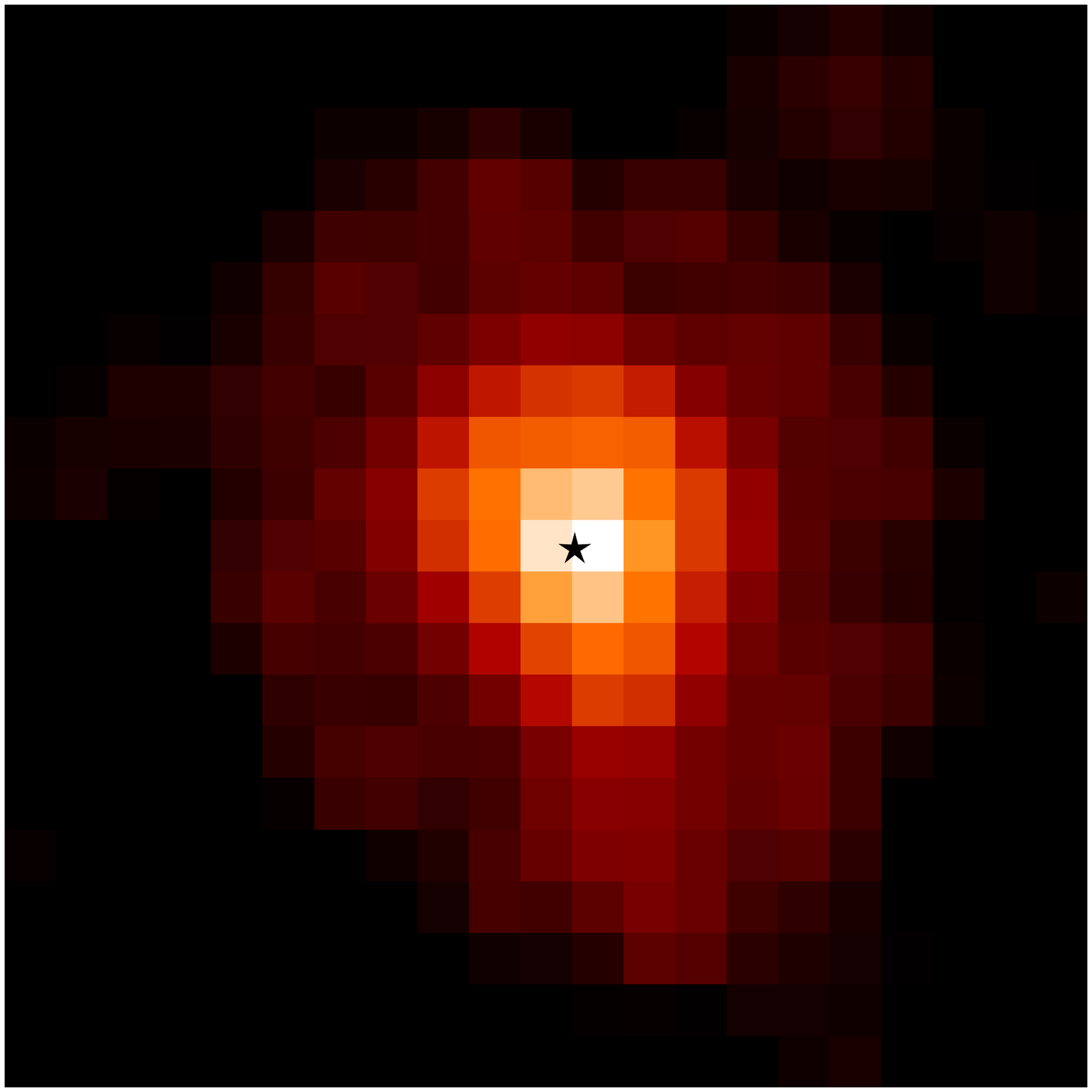}}
\caption{Postage stamps extracted from one of the two F435W HRC exposures for the two main components of HD 93129 
(A, left and B, center) and PSF used for fitting (right). A logarithmic scale between 0.1\% and 100\% of the peak 
value is used in all cases. The field size is $0\farcs53\times0\farcs53$. Star symbols are used to identify 
the positions obtained by PSF fitting.}
\label{pstamps1}
\end{figure}

\begin{figure}
\centerline{\includegraphics*[width=0.32\linewidth]{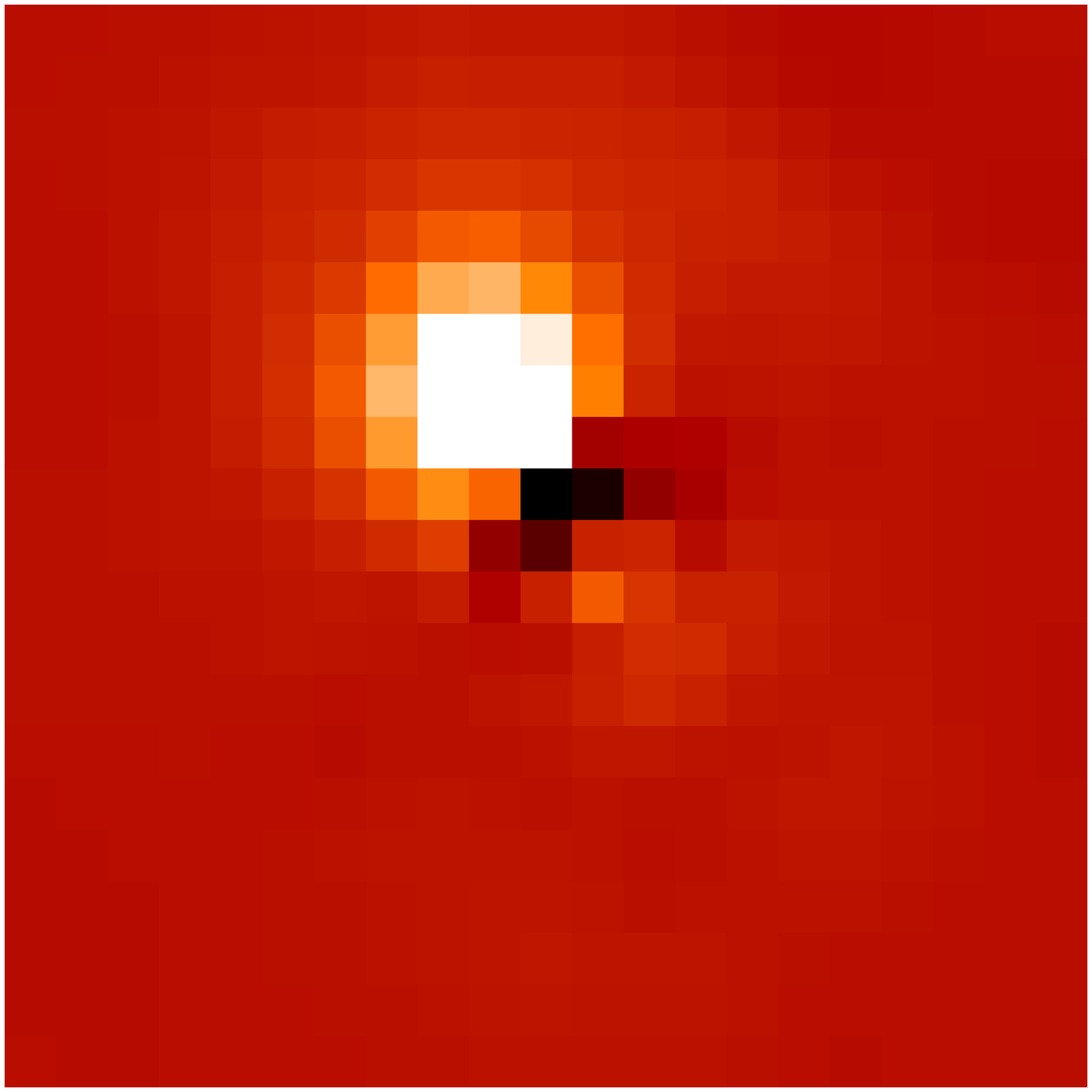} \
            \includegraphics*[width=0.32\linewidth]{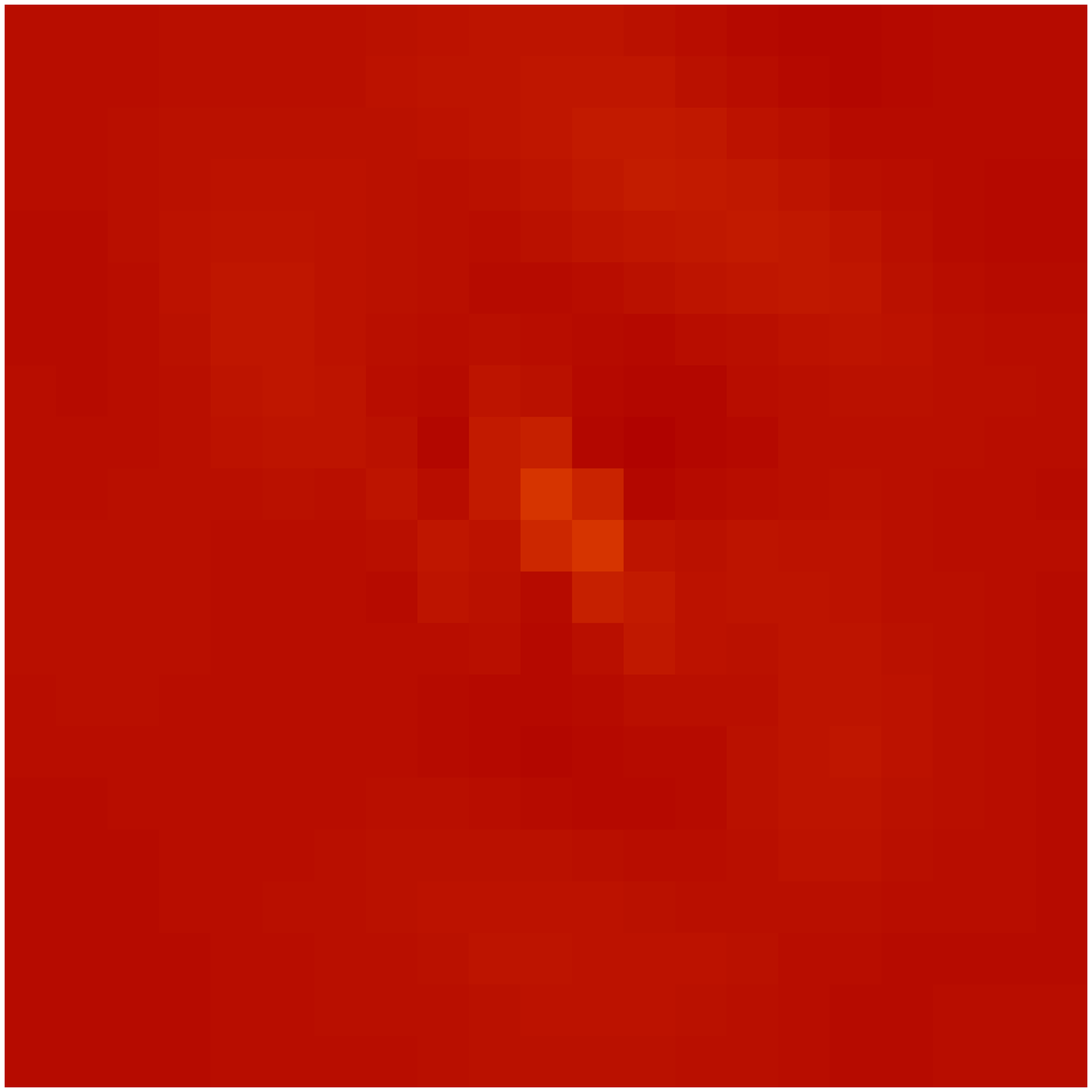} \
            \includegraphics*[width=0.32\linewidth]{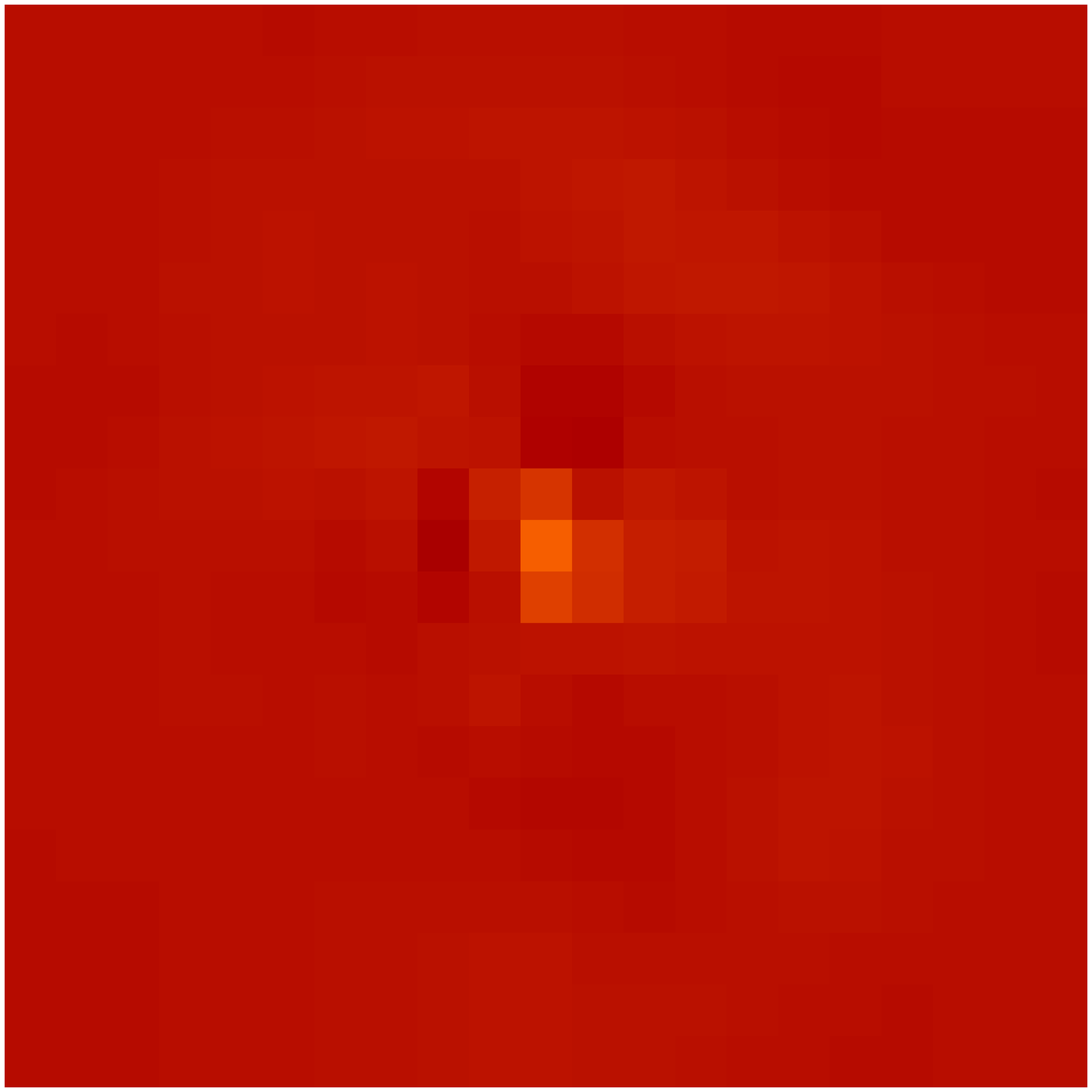}}
\caption{Fit residuals for HD 93129 A (left and center) and HD 93129 B (right) for one of the F435W exposures. The
central panel assumes two components (e.g. it treats A as a binary system) while the other two assume a single
component. A linear scale between $-3\%$ and $3\%$ of the peak value for the data is used in all cases. The residuals have
been smoothed with the PSF. The field size is $0\farcs53\times0\farcs53$.}
\label{pstamps2}
\end{figure}

	We applied a PSF-fitting IDL photometry code especially written for this purpose to the HRC data for HD 93129 A and B. 
The code was applied to 
two dithered exposures in each of the F220W and F435W filters, thus yielding four independent measurements for each star. If a single 
component is used for the fit, the residuals for HD 93129 B are very small but those of HD 93129 A are very large, as expected for a 
binary system (Figs.~\ref{pstamps1}and~\ref{pstamps2}). On the other hand, a two-component fit yields very small residuals for HD 93129 A, 
hence confirming the binary character detected with FGS. 

\begin{table}
\caption{PSF analysis for the two components of HD 93129 A.}
\centerline{\begin{tabular}{cccc}
 & & & \\
\hline
             & Separation (mas) & Position angle ($^{\rm o}$) & $\Delta m$                   \\
\hline
F220W exp. 1 & $50.4 \pm 2.5$   & $16.8 \pm 2.8$              & $ 1.1589 \pm 0.0087$         \\
F220W exp. 2 & $51.4 \pm 2.5$   & $13.3 \pm 2.7$              & $ 1.1751 \pm 0.0086$         \\
F435W exp. 1 & $52.0 \pm 2.5$   & $14.4 \pm 2.7$              & $ 1.1284 \pm 0.0055$         \\
F435W exp. 2 & $52.1 \pm 2.5$   & $13.2 \pm 2.7$              & $ 1.1244 \pm 0.0052$         \\
Final value  & $51.5 \pm 1.2$   & $14.4 \pm 1.4$              & $ 1.1670 \pm 0.0061$ (F220W) \\
             &                  &                             & $ 1.1263 \pm 0.0038$ (F435W) \\
\hline
\end{tabular}}
\label{hd93129}
\end{table}

	We present the results of the two-component PSF fitting to HD 93129 A in Table~\ref{hd93129}. The final row shows the proposed 
values for the separation, position angle, and $\Delta m$ derived from the four independent HRC exposures. Our position angle is 
18 $\pm$ 4 degrees to the E of that measured by \citet{Nelaetal04}, suggesting that we are detecting the relative motion of HD 93129 Ab 
with respect to Aa. From the 2.4 year difference between the epochs of the FGS and HRC observations we derive a very preliminary orbital 
period of $\sim$50 years for the system but, obviously, observations at other epochs will be needed to confirm and measure an orbit. We 
point out that this is the first O2/O3/O3.5 star ever measured to be an astrometric binary.

	The magnitude difference between HD 93129Aa and Ab measured from the HRC data is similar to but slightly larger than the one measured 
with FGS (which uses light with longer wavelengths). The two components have very similar F220W-F435W colors, with Ab being redder only by 
0.0407 $\pm$ 0.0072 magnitudes. Given their proximity, it appears unlikely that the relative color is caused by differences in the amount 
or type of extinction. On the other hand, such a difference in color is equivalent to that between 50 000 K and 44 000 K for TLUSTY models 
\citep{LanzHube03} with $\log g=4.75$ and solar metallicity, indicating that both components are likely to be early-O stars. We are 
currently working on a more detailed analysis of the photometry using CHORIZOS \citep{Maiz04c}.

	It is important to note that HD 93129 A has not been identified as a spectroscopic binary. With a separation between its two
components of about 150 AU, one would expect relative velocities of the order of 30 km s$^{-1}$ if the inclination is large. Therefore, 
its non-identification is not surprising, since one would require a separation an order of magnitude smaller to allow for a clear detection
of radial velocity variations\footnote{Of course, close binaries are easier to detect not only because the radial velocity variations are larger
but because they occur on shorter time scales.} (see e.g. \citealt{BonaStan05}). This also means that we 
cannot even discard the possibility that either HD 93129 Aa or Ab are binaries themselves.

	What does this mean for the biases in the top end of the IMF induced by multiplicity? Trumpler 14 is at an approximate distance of 
2.7 kpc. If it were located at the same distance as the Galactic Center or NGC 3603, HD 93129 A will likely appear unresolved with HRC or 
FGS. At the distance of the Magellanic Clouds, Aa and Ab would be unresolved and A and B could be resolved but only with HST or adaptive 
optics. Moving to M31 or M33, HD 93128 (another O3 star in the cluster outside the field in Fig.~\ref{trumpler14}) and HD 93129 would have 
an angular separation similar to that of Aa and Ab at its actual distance, with the rest of the stars (likely of late-O and B type) in 
Fig.~\ref{trumpler14} in between. This yields a total of (at least) four early-type stars blended together in a cluster that is quite
massive, but far less than R136. The reader can easily deduce from these simple calculations how much he/she can trust IMF derivations of
extragalactic young clusters that do not take into account multiplicity corrections.

\section{Conclusions}

$\,\!$\indent We have discussed three sources of biases in the determination of the initial mass function (IMF):
[1] the use of constant-size bins, [2] the uncertainty in the determinations of masses, and [3] the existence of unresolved multiple 
systems. Those three effects tend to produce IMFs that are flatter than the real one, all of them large enough to potentially introduce
significant systematic errors in the derived power-law slope. In the first case we present a technique that can get rid of the
bias almost completely, even after the data have been obtained. In the second case we present a method that can also be used a posteriori to
estimate the biases and we give some advice as to how to reduce them by using multifilter data. The third case, as demonstrated by the
example of HD 93129 A, is harder to correct, given that the current capabilities do not allow us to detect all multiple systems in the
stellar clusters and associations of interest, even those in our own Galaxy. Nevertheless, that should not preclude us from making
an effort to close the current gap between visual and spectroscopic binaries or from trying to estimate the contribution of 
of unresolved multiple systems to the IMF slope.

\acknowledgements             

Support for this work was provided by NASA through grants GO-09419.01, AR-09553.02, and 
GO-10205.01 from the Space Telescope Science Institute, Inc., under NASA contract NAS5-26555.


\bibliographystyle{aj}
\bibliography{general}


\end{document}